# The Innovative Behaviour of Software Engineers: Findings from a Pilot Case Study

Cleviton V. F. Monteiro
Universidade Federal Rural de Pernambuco
Rua Dom Manoel de Medeiros, s/n
Dois Irmãos, Recife-PE, Brazil
+55 81 33206590
cleviton@gmail.com

Fabio Q. B. da Silva
Universidade Federal de Pernambuco
Av. Jornalista Anibal Fernandes, s/n
Cidade Universitária, Recife-PE, Brazil
+55 81 21268430
fabio@cin.ufpe.br

Luiz Fernando Capretz
University of Western Ontario
Thompson Engineering Building
N6A 5B9, London, Canada
+1 519-661-2111
lcapretz@uwo.ca

## ABSTRACT
**Context**: In the workplace, some individuals engage in the voluntary and intentional generation, promotion, and realization of new ideas for the benefit of individual performance, group effectiveness, or the organization. The literature classifies this phenomenon as innovative behaviour. Despite its importance to the development of innovation, innovative behaviour has not been fully investigated in software engineering. **Objective**: To understand the factors that support or inhibit innovative behaviour in software engineering practice. **Method**: We conducted a pilot case study in a Canadian software company using interviews and observations as data collection techniques. Using qualitative analysis, we identified relevant factors and relationships not addressed by studies from other areas. **Results**: Individual innovative behaviour is influenced by individual attitudes and also by situational factors such as relationships in the workplace, organizational characteristics, and project type. We built a model to express the interacting effects of these factors. **Conclusions**: Innovative behaviour is dependent on individual and contextual factors. Our results contribute to relevant impacts on research and practice, and to topics that deserve further study.

## CCS Concepts
• Software and its engineering~Software development process management

## 1. INTRODUCTION
Innovative behaviour is a multidimensional construct defined as "*the intentional generation, promotion, and realization of new ideas within a work role, work group, or organization in order to benefit role performance, a group, or an organization*" [8]. Innovative behaviour is not the same as innovation. In order for innovation to happen, ideas must be generated, the best ones selected and implemented, and then deployed or marketed, generating profit for the organization. The first steps of innovation associated with the generation of new ideas, their promotion, and final realization in the workplace are the results of individuals expressing innovative behaviour. Examples of such behaviour include the suggestion of new products or processes, the adoption of new technologies, or the application of new working methods.

In our studies of industrial software engineering practice, we observed and catalogued several examples of the innovative behaviour exhibited by software engineers, with positive impacts at the individual, team, and organizational levels. For instance, during the investigation reported in this article, we observed a software engineer—responsible for the development and maintenance of an application database—implementing scripts to automate her manual work, thus both reducing errors and freeing her up to perform other productive activities. What is relevant in this example is that she voluntarily took the initiative to develop the script automations during her spare time (it was not among her duties or project tasks), and then promoted the new idea to her manager, finally incorporating the solution into the company's routine protocol.

The benefits of innovative behaviour in practice motivated us to investigate which factors foster or inhibit this behaviour at the individual, group, and organizational levels, in software engineering practice. Specifically, we were looking for answers to the following research question: *How is the innovative behaviour of software engineers supported or supressed in software development industrial practice?*

As a starting point, we conducted an *ad hoc* literature review covering innovative behaviour models from several fields. The findings showed incomplete and incomparable results, lack of established models or theory, and very few studies focused on software engineers and software organizations. Several authors have argued that pilot case studies are a suitable choice of research method to investigate a new phenomenon and to build theories when none are available or widely accepted [7][9].

Therefore, the goal of this article is to report the results of an industrial pilot case study developed to identify factors that influence the innovative behaviour of software engineers in practice. With these results, an initial model explaining the relationships among these factors was built, providing answers to our research question.

The remainder of this article is organized as follows. In Section 2, we present the background that supported this work. In Section 3, we describe the pilot case study design. In Section 4, we present the innovative behaviour model constructed with the results of the pilot case study. In Section 5, we discuss our results, their implications for research and practice, and present suggestions for future work. Finally, in Section 6, we present our conclusions.

## 2. BACKGROUND
In this section, the innovative behaviour construct will be detailed and compared to related concepts. We then summarize existing innovative behaviour models and aggregate their factors to build our initial conceptual framework.

### 2.1 Innovative Behaviour
The multidimensional aspect of the innovative behaviour construct comes from the definition of innovation that covers the proposal of new and useful ideas, their promotion, and their implementation [15]. Consequently, innovative behaviour is viewed as a multistage process that starts with an individual creating and proposing a new (potentially useful) idea. Then, this individual promotes the idea to gain support from colleagues, managers, or sponsors who can provide the resources necessary to help it materialize. Finally, the process culminates in the implementation of the idea, in the form of the production of a prototype, a proof, a concept, or the use of a new technology within a software project. Thus, different activities and different individual behaviours are essential at each stage [8][12].

Therefore, innovative behaviour differs from creativity because it is concerned with the promotion and implementation of ideas, while creativity only deals with the generation of new ideas [2]. It also differs from invention because those implemented ideas must generate value. Finally, it differs from innovation because it results from behaviour expressed by individuals, whereas innovation is the result of a process of idea generation leading to successful implementation, which generates value and/or profit. Such processes involve many other variables beyond the scope of an individual, such as the market, available resources, policies, strategy, etc.

## 2.2 Innovative Behaviour Models

The innovative behaviour phenomenon has been studied in fields such as health care [3], industrial corporations [15], knowledge-intensive service firms [12], and other industries [3]. Three existing models try to explain the antecedents of innovative behaviour, two at the individual level [3][15] and one studying the expression of this behaviour in the context of the working group [16].

In the model proposed by Åmo [3], individual innovative behaviour is positively influenced by twelve factors, which can be grouped into four categories:

- *The organization*: expressed strategy, and size of the organization;
- *The intersection between employee and employer*: position in the organizational hierarchy, organization's desires as expressed by management, culture of the work group, and level of specialization in job function;
- *The individual*: proactivity, intrapreneurial spirit, eagerness to learn, and age;
- *The innovation itself*: embedded learning potential and fitness with organizational goals.

Scott and Bruce [15] tested hypotheses relating individual innovative behaviour to factors in four categories: psychological climate for innovation; leadership; workgroup; individual characteristics of problem-solving style. Their findings in these four categories are summarized as follows:

- *Leadership* – two factors associated with leadership are significantly related to innovative behaviour: the quality of the leader-member exchange is related to the individual's perception of a climate as supportive of innovation; the leader-role expectation, i.e., the degree to which a supervisor expects the subordinate to behave innovatively, is directly related to innovative behaviour.
- *Climate for Innovation* – support for innovation is directly related to innovative behaviour because it creates the perception of a positive climate for innovation to take place.
- *Workgroup* – no significant relationship was found between the quality of team member exchanges and innovative behaviour. Similarly, team member exchanges are not related to the creation of a positive climate for innovation.
- *Individual problem solving styles* – a systematic problem solving style is negatively related to innovative behaviour, whereas no significant relationship was found with intuitive problem solving styles. Career stage is negatively related to innovative behaviour, meaning that individuals later in their career are less likely to behave innovatively.

Finally, West [16] proposes that the occurrence of group creativity and behaviour that moves toward implementation are influenced by a composition of four elements that interact with each other: group task characteristics, group knowledge diversity and skills, integrating group processes, and external demands. Particularly, West proposes external demand from the external environment or the organization itself as a new factor related to innovative behaviour. He contends that this relationship cannot be linear, but has an inverted U shape in the sense that too much or too little external pressure to innovate causes individual paralysis. In relation to group processes, this model describes effective conflict management, support for innovation, and the creation of intra-group safety as three factors linked to group creativity and innovation implementation.

## 2.3 Building an Initial Conceptual Framework

Analysing the three models briefly described above, we observed the following characteristics:

- Each model proposed different variables to explain innovative behaviour, with only a few overlaps.
- Some of the findings are potentially contradictory. For instance, the positive age relationship in Åmo's model and the negative career stage relationship in Scott and Bruce's model.
- Two of them [3][15] studied the innovative behaviour phenomenon at the individual level, while West [16] did it at the group level.

Although these characteristics make it difficult to relate or integrate propositions, it is still possible to group the factors into categories. Figure 1 illustrates a possible aggregation of the factors discussed in the three models, which are composed of seven categories directly related to individual innovative behaviour.

We used this synthesis as an initial conceptual framework in our case study. As defined by Merriam [13], the conceptual (or theoretical) framework is "the underlying structure, the scaffolding or frame of your study." As such, it is not an *a priori* theory from which hypotheses are derived to be tested. It offers a "system of concepts … that supports and informs your research" [13]. In particular, in our research was used to guide the construction of data collection instruments (namely, interview and observation scripts) and after data analysis was performed, it was used again to compare our findings with those of the existing models.

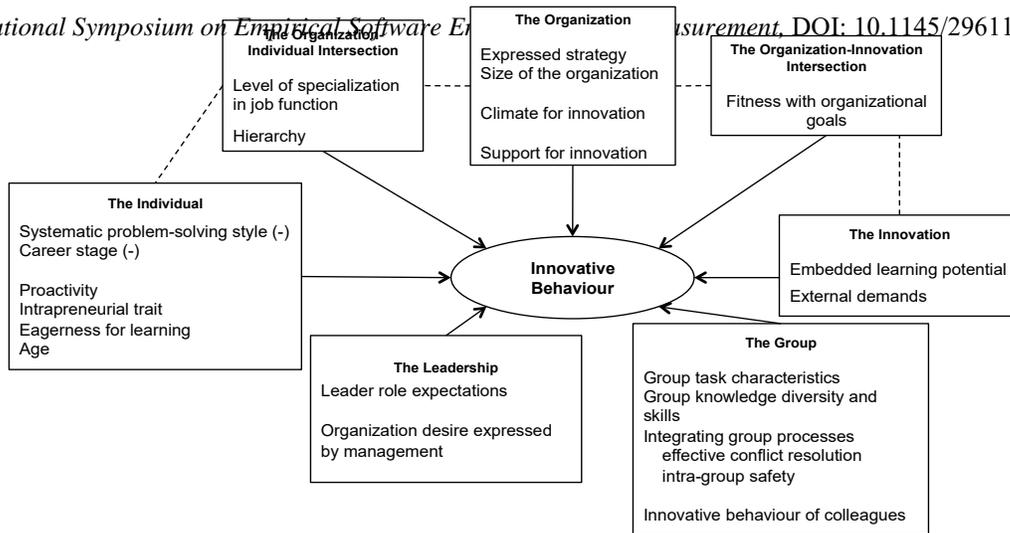

**Figure 1. Illustration of our Conceptual Framework**

## 3. METHOD

We are interested in understanding how individual software engineers interpret their experiences in the workplace regarding factors that potentially support or inhibit their innovative behaviour. Consistently with our interpretative/constructivist epistemological perspective, the nature of our research question, and investigated phenomenon, we performed a qualitative case study [13] and followed the method proposed by Eisenhardt [9] to build theories from case study research.

### 3.1 Getting Started

We started with a definition of the research question (Section 1) and the construction of the pilot case study design. The first and second authors worked together on construction of the case study protocol. The first author performed the data collection and analysis. The second author audited the data analysis, and together with the third author, reviewed the case study report. All three authors worked on the paper.

We chose the *software engineer professional* as the unit of analysis, supported by our conceptual framework and also because the research question is directly related to the expression of the phenomenon at the individual level. Further, the design also had to deal with contextual factors related to the unit of analysis. In this case, based on our conceptual framework (Figure 1), the following contextual aspects were considered:

- The Group: team influence on the individual was considered.
- The Leadership: in the team, the leaders exert different type of influence on individuals than other team members.
- The Organization: the organization has cultural aspects, an organizational structure, norms, and values that might influence the behaviour of individuals.
- The Innovation: the nature of the innovation itself, and its relationship with the organization and individuals within it.

We created a flexible design to allow for the exploration of the phenomenon of innovative behaviour, the identification of relevant variables, and their relationships (Figure 2). We investigated a single software organization, so that organizational factors would be as similar as possible across projects. We then studied different individuals from two different projects, each one with different team leaders, to maximize the diversity and richness of the collected data. To obtain variability regarding the individuals who participated in the study, we used the criterion detailed in Section 3.2.

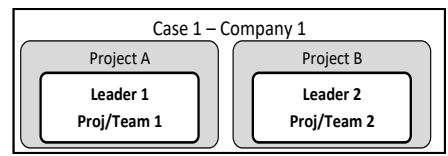

**Figure 2 - Pilot Case Study**

### 3.2 Selecting participants

We were interested in selecting individuals with low, medium, and high levels of innovative behaviour to be able to compare the behaviours and what influenced them. Therefore, to select the participants, the project manager of each project classified the team members according to the frequency with which they behaved innovatively, following a definition of innovative behaviour that was explained in person by the researcher. Then, based on the managers' classifications, the researchers ranked the individuals and chose members with low, mid-rank, and high frequency innovative behaviour from each team. The project managers then confirmed the researchers' assessments and choice of participants. The project managers were also interviewed to allow for data triangulation.

### 3.3 Collecting Data

We used more than one source of data and method of collection to increase consistency and reliability [13]: interviews and field observations.

Semi-structured interviews were performed with software development team members and their leaders (project managers). The interview script for team members contained an overview of the study used to inform the participants along with 77 open-ended questions. The questions covered: team members' backgrounds, their innovative behaviour, organizational context, working group, leadership, and the individual's characteristics. The interview script for team leaders had a similar structure, including 27 questions grouped into the following categories: leaders' backgrounds, organizational context, working group, leader, subordinates, and again, the individual's characteristics. The subordinate questions were asked regarding each member that participated in our study. Both interview guides were composed of open questions combined with probing questions. We used pilot interviews to test and refine the interview guides. These were performed with individuals from another company that did not participate in the study.

Four team members and two project managers were interviewed. The interviews were recorded and comprised 5 hours and 2 minutes of audio.

Observation was chosen to allow the researcher to observe behaviours and interactions among team members that could not be obtained from interviews [13]. Combined with the interviews, this type of data collection allowed for data triangulation, which improves the reliability of results [9][13]. The observations happened during project meetings because it was one of the only times when individuals interacted face-to-face during the project. The observer (first author) took notes focused on identifying idea proposals, and past or present implementation of an idea proposed by team members, as well as anything that referred to the past existence of such behaviour that could be further explored after the meeting. Two meetings about each project were observed.

### 3.4 Analysing Data

Data analysis was performed in tandem with data collection, in incremental and iterative steps. We used qualitative coding techniques to code, categorize, and synthesize data [13]. All audio from interviews was transcribed verbatim. We used QSR NVivo[1] to support the data analysis and synthesis. Data analysis began with open coding of the transcripts. Post-formed codes were constructed as the coding progressed, and were attached to particular pieces of the text with the support of NVivo. The following scheme was used to trace the evidence from the data.

<company code><Project code><Individual position><Individual code>_<open code>

An example of a complete code is *C1PATM1_Intrinsic_need_to_try_new_ideas*, which means that the evidence points to the code "Intrinsic need to try new ideas" collected from the interview of team member 1, which worked on Project A, in Company 1. This code was attached to the following excerpt.

> *"For me, doing the same monotonous thing is hard for me to do because it becomes boring. So I have this intrinsic need to try to make things... The problem solving, the discovery, etc."*

Codes arising from each interview were consistently compared to other codes in the same interview and from other interviews, and to data from the observations. The constant comparisons of the codes helped us group them into categories of factors that were related to innovative behaviour. In the example above, we grouped the code *C1PATM1_Intrinsic_need_to_try_new_ideas* under the category "Individual Observable Signs and Behaviours." As the process of data analysis progressed, we built up the interacting effects of these factors and created a model that describes the innovative behaviour of individuals in this organization.

Finally, data triangulation was performed comparing the codes (and content) extracted from the individuals' interviews with data from the leaders' interviews, as well as with the observation notes. For example, the behaviour explained above by the participant C1PATM1 could be confirmed with the following excerpts extracted from her manager's interview.

> *"She is one of the more outgoing people on the team. She pass, send her ideas ... I would say she is a kind of extreme side of spectrum, you know, send me her ideas, sometimes too many ideas coming out and we can't do everything at the same time."* [C1PATL]

In addition, the observation notes were used in the triangulation.

> Observation note: *"C1PATM1 searched for a new solution to solve the problem of the rich text component and proposed it in the meeting (this task was not assigned to her). This component is being used in many places, but the current solution is not working as it should be."*

### 3.5 Enfolding the Literature

Following Eisenhardt's guidelines [9], after completing the data analysis and shaping an initial version of the model, we looked at the literature to sharpen our definitions of constructs and to raise the theoretical level. To do this, we compared the resulting model with those presented in Section 2, using the structure of our conceptual framework to guide the comparisons.

### 3.6 Reaching Closure

In qualitative research, when to stop collecting data or sampling new participants is an important decision. According to Charmaz [6], the standard answer is to stop when theoretical saturation has been achieved. However, according to the author, theoretical saturation is misleading because it is difficult to prove and can be achieved by superficial analysis of the data. Therefore, Charmaz [6] advocates that instead of basing our decision on theoretical saturation, we should guarantee that categories are consistently built from the data, i.e., we should look for theoretical sufficiency instead of saturation.

Our data analysis was performed by one researcher and thoroughly reviewed by the other two. An audit trail was generated and multiple sources of data were consulted for triangulation purposes. According to Merriam [13], these procedures increase confidence in the consistency and thus, the theoretical sufficiency of the findings. Further, we sampled a diversity of individuals, providing for richer data and more expressive results.

### 3.7 Ethics

The company signed a Term of Authorization and the researchers signed a Non-disclosure Agreement (covering access to sensitive information). Each participant signed an Informed Consent Form that explained the overall objective and relevance of the research, guaranteed data confidentiality, the anonymity of participation, the non-obligatory nature of participation, and the right to withdraw from the research at any time. All invited individuals freely agreed to participate and no participant withdrew from the research. After data analysis, we asked the participants for explicit permission to use the quotes. They all agreed to the use of all quotes presented in this article.

## 4. RESULTS

We start with a description of the research context and then present the results of the pilot case study.

### 4.1 Context Description

This section describes the context of this research: the software company, the selected projects, and the participants.

#### 4.1.1 The Software Company

The pilot case study was performed between November 2012 and July 2013 in a software development company that specializes in customized software outsourcing as well as business intelligence (BI) services, hereafter called the Company. The Company is based in Toronto (Canada) and was founded in 1994. During the case study, the Company had 45 professionals, ranging from 30 to

---

[1] www.qsrinternational.com/products_nvivo.aspx

48 years old, from different ethnic backgrounds. These professionals were designers, system administrators, system analysts, software engineers, software testers, BI specialists, database administrators, project managers, and a human resources manager.

The three company owners directed the Company and the Company's professionals were employees or contractors. The organizational structure was flat, with managers reporting directly to the directors/owners. In some projects, the directors were involved in certain decisionmaking together with the software development team. The Company's projects explored several areas, including e-Health, energy and environment, financial services, media, etc.

*4.1.2 The Projects and Participants*

As discussed in Section 3.2, we selected participants from two projects to achieve variation regarding innovative behaviour. We present the aggregated profile of participants in Table 1.

In Project A, the team was composed of 15 members: one project manager, two business analysts, one quality analyst, and ten software engineers. Two of these developers were also software architects. They were developing a web system for the health insurance area. At the time of the interviews, the system was being developed for the purpose of substituting a legacy system and was being designed to achieve close to the same functionality and workflow. The manager reported that the process followed by the team was based on Scrum. Three members of this project were selected to participate, in addition to the project manager (see Table 1).

In Project B, the team was composed of nine people: one project manager, one technical leader, six software engineers, and one business analyst/tester. They were developing a new web decision support tool for a health insurance company based in the USA. The manager reported that the process followed by the team was based on Scrum. In addition to the project manager, one member of this team with high innovative behaviour participated in the interviews.

## 4.2 Factors Related to Innovative Behaviour

We identified individual characteristics and work-related factors that were related to innovative behaviour according to the participants' perceptions. All factors were identified during open coding and their precise definition evolved throughout the process of data analysis and comparison with literature. With the help of axial coding, the identified factors were grouped into categories and the relationships among them were expressed as hypotheses. Finally, we built a model representing these relationships that explains the expression of innovative behaviour by software engineers in this case study.

When presenting the results, we use excerpts from interviews as supporting evidence to build internal validity. We use the code *HIB* with quotes to denote individuals with high innovative behaviour (confirming with their managers and also based on our observations), *MIB* for those with medium innovative behaviour, and *LIB* for those with low innovative behaviour.

Participants expressed different **Individual Attitudes** towards innovative behaviour. Those individuals with high innovative behaviour valued new ideas and experiences in the workplace more than those that presented low innovative behaviour, as expressed in the following quotes:

*HIB*: *"For me doing the same monotonous thing is hard for me to do because it becomes boring. So I have this intrinsic need to try to make things… The problem solving, the discovery, etc."* [C1PATM1]

*LIB*: *"I'm not a theory person, … I would have not an incentive to research a new idea or new way to do things. But they are totally personal things."* [C1PATM3]

We identified external signs or behaviours expressed by individuals that had this positive attitude. They were more *open to new experiences*, *curious*, *proactive when it came to identifying problems*, *liked to learn*, and *were often looking for new technologies*. They possessed these behaviours even in the presence of situational factors that could inhibit innovation, such as the change-avoiding attitudes of colleagues or poor leadership feedback:

*HIB*: *"I had a lot of ideas and I tried to push through a couple of things… I did a couple of experiments because of one idea I had. I wanna do this because I would streamline all of the processes of the whole company."* [C1PATM1]

On the other hand, those with low innovative behaviour would prefer following familiar processes, best practices or known technologies. They would show less curiosity and less proactivity in identifying and solving problems.

*LIB*: *"we don't do pure science here, so we are not based on a theory to find a practical solution for that. Basically we are based on requirements and we need to use system best practices to meet the client requirements."* [C1PATM3]

The above findings lead to the first hypothesis:

**Hypothesis 1** – *The* **positive individual attitude** *regarding the proposition of new ideas, their promotion, and implementation in the workplace will directly contribute to the individual's expression of* **innovative behaviour**.

However, the transformation of positive attitudes into idea generation, promotion, and realization (i.e., innovative behaviour) was contingent on situational factors. Even those individuals with positive attitudes would change their behaviour when they were confronted by repetitive rejections and perceived, through direct or indirect *feedback*, that proposing ideas and implementing them was worthless for colleagues, leaders, or the company in general. The following excerpt is about a company in which the participant worked before.

*HIB*: *"… [But] even though it was a good idea, no one else wanted to do it. Because they have done things in a certain way and these people didn't care about working together as a team, … And eventually I got to the point that I really stopped to propose ideas…"* [C1PATM1]

This leads to a second hypothesis about the effects of *feedback* (from peers, team leaders, and the organization) on a previous innovative behaviour, and on future expressions of this behaviour.

**Hypothesis 2** – *The* **feedback** *on the expression of a past innovative behaviour will indirectly influence the expression of future* **innovative behaviour** *through its moderating effect on the relationship between the individual attitude and her innovative behaviour.*

Further analysis of our data revealed that, apart from feedback, other types of peers or colleague behaviour in the workplace influence the expression of an individual's innovative behaviour. Regarding that behaviour, we were able to further distinguish the influence of two categories of factors: those related to team leaders and those related to other team members. We chose to treat those factors separately because leader influence was different from the type of influence exerted by other team members. In our study, the leaders were the project managers and they were responsible for schedule, budget, and scope management, and also for contact and negotiation with clients.

Thus, they had more power to promote ideas and secure resources, as well as time to implement them.

The influence of **Leader Behaviour** was related to two complementary factors. First, the perception of individuals regarding the leader's willingness to accept new ideas (*idea acceptance*) would stimulate idea proposal. In addition, the individuals were also stimulated when the leader promoted the ideas proposed and acquired resources for their implementation (*idea championing*).

> LIB: "I think the innovations require support from your manager. Because when you want to innovate you need to invest some of your time. Sometimes a new equipment ..." [C1PATM3]

On the other hand, when a leader did not accept new ideas, the individual would stop expressing innovative behaviour. The answer below was provided when a member was asked about aspects that did not stimulate her to behave innovatively in the company.

> LIB: "I don't see it here... I had a project before [on another company] that the manager was like a dictatorship, kind of. Sometimes he had a certain way to achieve such task. But it may not be an efficient way or the best way from the company point of view...the way he will address the situation is by authority. So there is some conflict. The way we resolve here in Company is discussing it. And try to find a middle ground." [C1PATM3]

Two types of **Team Member Behaviour** could also have had moderating effects on this particular individual's attitudes and her innovative behaviour: *idea acceptance* and *conflict resolution*. The former is similar to a leader's influence because behaviour that promotes change in colleagues and the corresponding feedback provided would inhibit the individual when these ideas were not accepted, or stimulate her to share ideas and promote their implementation when they were accepted.

> HIB: "I think if I work with people here at [Company] who haven't any interest to listen to ideas, then I will not propose anymore [to these people]" [C1PBTM1]

In addition, the way the conflicts were resolved was important because when there was space for discussion and decisions were shared, the individuals perceived that they had a voice and they were not inhibited due to authoritarian decisionmaking or colleagues' disagreements.

> HIB: "If I felt strange about something and my co-worker felt strange about something else, we will certainly argument by figuring out what of those is best to our customer... New ideas come up from that? I think so." [C1PBTM1]

In addition, debates regarding the idea or the conflicting aspects, if conducted effectively, could generate new ideas or better ways of implementing ideas. Therefore, good or effective conflict resolution at the team level creates a positive environment for innovative behaviour.

Situational factors related to peers, leaders or team members, will influence the individual's innovative behaviour, moderating the relationship between the individual attitude and her behaviour. We expressed these influences with the following hypotheses.

**Hypothesis 3** – *The perception of the individual about peers' (team members or leaders) idea acceptance will indirectly influence innovative behaviour through its moderating effect on the relationship between individual attitude and the expression of innovative behaviour.*

**Hypothesis 4** – *The perception of the individual about leaders' idea championing will indirectly influence innovative behaviour through its moderating effect on the relationship between individual attitude and the expression of innovative behaviour.*

**Hypothesis 5** – *Effective conflict resolution at the team level will indirectly influence innovative behaviour through its moderating effect on the relationship between individual attitude and the expression of innovative behaviour.*

Although we did not investigate organizational climate in this study, based on the literature on conflict management [11] we believe that Hypothesis 5 could be refined as follows:

**Hypothesis 5r** – *Effective conflict resolution at the team level will indirectly moderate the relationship between individual attitude and the expression of innovative behaviour, due to the establishment of a team climate that is supportive of new ideas and change.*

At the organizational level, we identified two potential moderators in the category **Organization's Context**: *bureaucracy* and the *support for innovation*. Bureaucracy has a negative moderation effect on promoting change, because the more difficult it is to get ideas approved and the slower the process of implementing change, the less innovative behaviour individuals exhibit.

> LIB: "...[another company I worked] may not be dynamic enough to adapt to the changes that our client face all the time... everything is rigid, in the sense of rigid procedure... so sometime they may have a sense of [only accept change] when the company policies come in. So they will be a little bit more passive in adopting it. But here [at Company] is a bit different." [C1PATM3]

Company support was important for individuals to feel comfortable expressing innovative behaviour. If the company provided resources, for example, or time to implement an idea, the individual would not have to spend extra effort just to try something new. Company support created a feeling of belonging in the company and a climate perceived as supportive for innovation which, in turn, increased individual commitment to innovative tasks.

> MIB: "So if personal initiative start to make the innovations or trigger the innovation, [then] if the company's support it to your work you will have a sense of belonging to the company. ... you will feel more belonging to the company and you will feel more commitment to have the project or to have the task completed." [C1PATM2]

These findings lead to the following hypothesis:

**Hypothesis 6** – *Organizational bureaucracy will indirectly influence innovative behaviour, through its moderating effect on the relationship between individual attitude and individual innovative behaviour, and this effect will be negative.*

**Hypothesis 7** – *Organizational support for innovation will indirectly influence innovative behaviour, through its moderating effect on the relationship between individual attitude and individual innovative behaviour.*

Participants also perceived that the software **Project/Task Type** or some of its characteristics would affect opportunities to express innovative behaviour. In particular, two project characteristics were explicitly identified: *requirements stability* and the *technical challenges* to implement these requirements. Both shaped the type of ideas that the individual could propose and limited the type of resources the engineers could ask for. The requirements stability had a negative moderating effect, because the more stable the requirements were—in the sense that they could not be changed or there was no incentive to do it—the less innovative behaviour individuals would exhibit. For example, projects with predefined requirements, such as Project A, did not have space for new requirements because the new systems had to provide the same functionality as the previous one:

> MIB: "For this particular task I'm working on, not really [have to be innovative]. My job right now is to make sure that when we migrate from one platform to another we don't lose stuff. We should maintain *consistency. We properly document things and we properly test things." [C1PATM2]*

Therefore, the individuals were constrained then, and their ideas used to be more related to the development process and technology adoption rather than on new products or new requirements.

In turn, the *technological challenges* had a positive moderating effect, because when there were few or no challenges, the individual perceived fewer opportunities to implement new solutions, i.e., they would replicate already existing solutions and technologies to deal the problem at hand. Thus, the innovation expectancy in these cases was lower and individuals perceived less space in which to innovate.

*HIB: "The projects we work on... they are a little dry. It is basically boring enterprise stuff. Business database type, data mining. We are not really pushing development as far as technology goes. So in that respect is that... ok... it is not exciting." [C1PATM1]*

Conversely, when challenges were faced, they would perceive more space in which to solve the problem or implement the solution proposed, thus expressing more innovative behaviour.

These findings lead to the following hypotheses:

*Hypothesis 8 – The technological challenge of the project and its tasks will indirectly influence innovative behaviour, through its moderating effect on the relationship between individual attitude and individual innovative behaviour.*

*Hypothesis 9 – The requirement stability of the project will indirectly influence innovative behaviour, through its moderating effect on the relationship between individual attitude and individual innovative behaviour, and this effect will be negative.*

Finally, we observed that some individuals were naturally motivated to behave innovatively and this behaviour had occurred when they worked in different companies, demonstrating that certain non-contextual factors also affect innovative behaviour. Further, even when situational factors imposed constraints that could potentially inhibit their innovative behaviour, they still behaved innovatively at least for some time. In the following excerpts we give examples of different individual perceptions of the rejection of their ideas by their colleagues. In the first example, the participant from Project A said that she tried to push her ideas for some time even facing difficulties in the company:

*HIB: "If I think that something should be done then I will push it. Sometimes maybe a little bit too much. So lot of the times I usually got my way in what need to be done, even with the company being a bit slow." [C1PATM1]*

On the other hand, some individuals were inhibited immediately by the rejection of their first ideas as exemplified in the following excerpt from a member of Project B.

*HIB: "When you propose an idea, if the person is receptive there is no problem. If the person shut it down, and say your idea does not make any sense, that is what stop you." [C1PBTM1]*

These behaviours cannot be completely explained by the moderating effects of the situational factors. Different individuals react differently to the same behaviours by their peers and to the same organizational support, for instance. To explain these behaviours, we hypothesize that the moderating effects of these situational factors are also moderated by individual personality traits.

*Hypothesis 10 – Individual personality traits will directly affect the strength of the influence of situational factors on innovative behaviour through its moderating influence on the relationships between situational factors and the expression of individual innovative behaviour.*

In this study, we did not investigate personality directly. Therefore, this hypothesis needs to be refined and tested in future studies by identifying which personality traits are important and how this moderating effect works.

## 4.3 A Model of Innovative Behaviour

We integrated the findings to create a model that represents the relationships expressed in the hypotheses raised during our data analysis. The model, called Initial Innovative Behaviour Model for Software (IBMSW-*i*), is depicted in Figure 3.

According to IBMSW-*i*, the individual attitude towards proposing, promoting, and implementing new ideas is directly related to the expression of innovative behaviour (Hypothesis 1). We also observed that individuals with positive attitude towards innovative behaviour would show signs and exhibit certain behaviours such as curiosity, proactive problem identification, a desire to learn, openness to new experiences and they were often looking for new technologies.

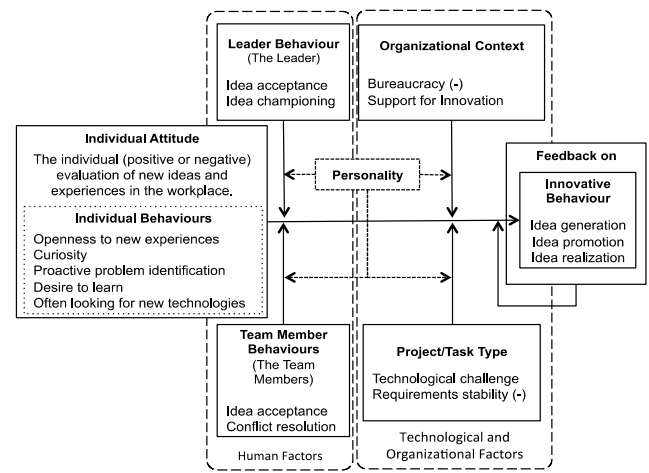

**Figure 3. The IBMSW-*i***

The expression of innovative behaviour is influenced by situational or contextual factors in the workplace. These factors create workplace conditions that will be perceived and interpreted by individuals, and will in turn moderate the expression of innovative behaviour at the individual level. If individuals perceive that the workplace has favourable conditions, they will be stimulated to express their innovative behaviour. A workplace perceived as non-favourable would tend, in turn, to supress the expression of this behaviour. We can group these categories of factors into two higher-level categories: those containing **Human Factors**, and those containing **Technological and Organizational factors**.

In the higher-level category of **Human Factors**, two categories were associated with the observable signs or behaviours of working peers or colleagues. In this case, idea acceptance creates favourable conditions for idea proposition (Hypothesis 3), whereas idea championing and effective conflict resolution create further favourable conditions for idea promotion and implementation (Hypothesis 4 and 5, respectively). In particular, effective conflict resolution seems to create an organizational and/or team climate in which individuals are more willing to expose their ideas (refined Hypothesis 5r). From these findings, we can build a hypothesis:

*Hypothesis 11 – The relationship with peers (team members and leaders) at the workplace will indirectly affect the*

*expression of innovative behaviour through the creation of (favourable or unfavourable) working conditions for idea proposition, promotion, and implementation.*

The organization as a whole also influences the expression of innovative behaviour. Bureaucracy and support for innovation have opposite influences, with bureaucracy being negatively related and support for innovation positively related to the expression of innovative behaviour (Hypothesis 6 and 7, respectively). Project and task types will also influence the expression of innovative behaviour through the level of requirement stability and the technological challenges associated with the tasks (Hypothesis 8 and 9). This is related to task uncertainty at a more general level. The organizational factors and the uncertainty levels of the tasks are likely to be interrelated, as expressed in this hypothesis:

**Hypothesis 12** – *Higher levels of task uncertainty (requirements flexibility and technological challenge) in the presence of support for innovation and low bureaucracy in the organization will indirectly affect innovative behaviour, through its moderating effect on the relationship between individual attitudes and individual innovative behaviour.*

The relationships expressed in Hypothesis 11 and 12 are moderated by individual personality, as expressed by Hypothesis 10. We postulated in Hypothesis 10 that individuals would react differently to the situational factors depending on their personality traits.

Finally, the expression of innovative behaviour evolves over time, contingent on the feedback received. Positive feedback on past innovative behaviour is likely to stimulate the individual to continue expressing this behaviour. Conversely, negative (or an absence of) feedback will adversely affect the future expression of innovative behaviour (Hypothesis 10).

## 4.4 Enfolding Literature

As the final step in our data analysis, we compared our findings with the models discussed in Section 2.2. We structured this comparison according to the five categories in our model.

At the *individual level*, the most important distinction between our understanding of the phenomenon and the understanding shown by other models is regarding the role of individual attitude. Consistently with the literature on organizational psychology [1], we understand that individual attitude drives the individual towards expressing or not expressing some behaviour. The other models characterized other observable signs or behaviours as antecedents or potential predictors of innovative behaviour. We also found some of those antecedents, such as *proactivity* and *eagerness for learning*. However, we contend that those behaviours and innovative behaviour are originate from the individual attitude towards change and innovation. Åmo [3] and West [16] discussed individual attitudes in their work. Åmo discussed that certain identified traits of innovative individuals may be related to individual attitudes, but did not express this directly in his model. West only addressed attitude change and the role of change in innovation.

With respect to the behaviour of the *leaders*, we found that idea acceptance and idea championing are indirectly related to innovative behaviour through their moderating effect on the relationship between individual attitudes and the expression of innovative behaviour. This is consistent with the findings of Scott and Bruce [15] in which the quality of leader-member exchange was related to the creation of a positive climate for innovation as well as to the innovative behaviour itself. At the *group level*, we found that idea acceptance and effective conflict management also indirectly affected innovative behaviour. Similarly, West [16] also identified effective conflict resolution and the creation of intra-group safety as important to the expression of innovative behaviour at the workgroup level. These findings seem to indicate that the quality of workplace relationships may create a psychological climate that will be perceived as supportive of innovations. Our Hypothesis 5r is in agreement with this interpretation.

At the *organizational level*, the embedded learning potential from Åmo's [3] model is closely related to our findings regarding technological challenges, because such challenges are likely to offer learning potential, although technological challenge entails other aspects. The characterization of external demands in West's [16] model is related to technological challenge and also to requirement stability; demands with more technological challenge and less stable requirements will offer more space for innovative behaviour. We have not investigated the curvilinear relationship between these factors and innovative behaviour as proposed by West, but it seems plausible that too much challenge and requirement instability would result in a cognitive overload and negatively affect the expression of innovative behaviour.

We did not observe the effects of organizational factors related to the expression of an organization's strategy, the fitness between strategy and innovation, the hierarchical position of an employee, or the size or the organization. This is because in our case we were dealing with a small company in which the communication of strategy was informal and the hierarchy was flat. The aforementioned factors are likely to be more relevant in larger companies. Similarly, we did not observe the effect of age or career stage because we sampled participants with similar characteristics in these aspects. This was a case-specific limitation, because the company employees are all of similar age and were at similar stages in their career. These limitations should be addressed in future studies.

Finally, our characterization of the influence of *feedback* on future expressions of innovative behaviour is novel. Although Scott and Bruce [15] address the quality of leader-member interactions, they did not explicitly conduct studied feedback. Further, the moderating role of personality proposed in our model is also novel.

A comparison to the existing models gives more strength to our findings. First, some factors we identified in the software engineering area are consistent with those presented in the literature. Second, we identified factors like requirement stability and technological challenge, which refined the general notion of external demands proposed by West [16] with the specific characteristics of software development. Third, we explicitly expressed individual attitudes as a key driver for the expression of innovative behaviour, which has not been considered in the other models. Finally, we identified the explicit role of feedback in innovative behaviour, also a novel finding.

## 5. DISCUSSIONS

In this section, we discuss the limitations and validity of our results, and the implications of these results for practitioners and researchers, showing directions for future research.

## 5.1 Addressing Limitations, Validity and Reliability

Validity and reliability assessments used in positivist experimental studies do not apply directly to interpretive qualitative research. We discuss the validity and reliability of our results from the perspective proposed by Merriam [13].

*Construct* validity in qualitative research is related to the precise and clear-cut definition of constructs that is consistent with the meanings assigned by research participants. We compared and contrasted the definitions interpreted from our data with the literature. Whenever the meaning assigned by the participants differed from the literature, we double-checked with the participants until a consistent definition was reached.

Internal validity, or *credibility*, refers to the extent to which the results match reality and that the researchers were able to capture reality as closely as possible. To increase credibility, we tried to achieve maximum variation on the sources of data. We collected direct data from participants in different teams and with different levels of expressed innovative behaviour. We also contrasted and compared interview data with observations. Finally, we compared the findings with the literature to sharpen construct definition and increase internal validity. One limitation regarding the sampling of participants is that we used a subjective assessment given by the team's manager about the individuals' innovative behaviour. Subjectivity in this assessment could have had an impact on internal validity. Thus, considering this limitation, a data analysis was performed and observational data was used to improve internal validity.

In qualitative research, we should strive to build results that are transferable (instead of generalizable in the positivist sense). Therefore, although we do not expect all our findings to be directly applicable to other contexts, it is possible to learn from the case description and decide to what extent the findings can be applied or transferred to other situations. Two strategies were employed to enhance the *transferability* of the results. First, we tried to provide a rich description of the research method, the context in which the research was performed, and the results themselves, although we believe this is one of the limitations of this article, since space restriction impacts the possibility of rich and detailed descriptions. Second, we sampled the participants to achieve maximum variation because this would provide richer data and, consequently, a more comprehensive and widely applicable model.

Finally, one potential criticism about this study is that it is a small scale, pilot study involving few participants. In fact, this was a design decision because the phenomenon had not been investigated in software engineering before and the existing literature from other areas was not conclusive about which factors should be observed and analysed. Christie et al. [7] suggested the use of pilot case studies in such a context. We then opted to perform a low cost and relatively fast pilot study to explore the phenomenon, create a preliminary model, and identify relevant research variables that could guide the design of a more comprehensive, full-scale case study design. We believe that our results achieved this goal. Further, we also believe that our results have important implications for practice and research, as we will explain in the following sections.

## 5.2 Implications for Practice

Our model shows that individuals will express innovative behaviour depending on their individual attitudes, moderated by the existence of favourable contextual conditions. In a psychological climate perceived as supportive of innovations, individuals with a positive attitude will tend to express more innovative behaviour than those with a negative one. However, the levels of uncertainty related to technological challenge and requirements stability of projects/tasks also influence the expression of innovative behaviour. Projects with higher uncertainty will offer more space for new ideas, and therefore, stimulate the expression of innovative behaviour on individuals. On the other hand, stable projects with fewer challenges or uncertainties are not likely to support innovative behaviour. Practitioners should be aware of these findings because it is unlikely that innovative behaviour will be expressed under unfavourable conditions.

Further, according to contemporary studies about work motivation [10], organizations should match different individuals' needs and desires to the types of tasks they perform. Therefore, software organizations should try to match an individual's desire or interest in expressing innovative behaviour and the conditions that must be available in the workplace for this behaviour to actually be expressed. Our model can be used to guide this matching. First, it identifies signs or behaviours of individuals that are likely to express innovative behaviour. These observable signs can be used to identify individuals that would naturally behave innovatively, given the right contextual conditions. Tasks or projects with higher technological challenge and less stable requirements could be allocated to such individuals.

Although most of the time the level of uncertainty of projects is defined by market or other organization-wide factors that may be difficult to manage, the factors related to the leaders and team members' behaviours are less difficult to change. In particular, we identified that feedback on previous innovative behaviour can have a significant impact (positive or negative) on future innovative behaviour. In an environment in which ideas are valued and conflicts are effectively managed, timely and consistent feedback is likely to foster the continuous flow of ideas in the organization.

## 5.3 Implications for Research

Although this pilot case study provided solid results that improved our understanding of the phenomenon of innovative behaviour in software engineering, future research will certainly extend and improve our results. Here we summarize the issues that arise from this pilot study.

*Moderating the effect of personality*: in our model we postulated that individual personality would moderate the strength of the external or situation factors on innovative behaviour. Future research could study these effects, identifying how individuals with different personality traits react to external factors related to the team, the organization, and the external environment in general.

*Project/task uncertainty*: different task and project characteristics will offer more or less space for innovation and, thus, to the expression of innovative behaviour. We identified that technological challenges and stability of requirements are two project characteristics related to innovative behaviour. We propose that research on this topic should look at project or task uncertainty in a more general way to identify other potentially relevant factors.

*Leadership style*: we identified that certain leadership behaviours were important for creating a psychological climate perceived as positive for innovation. Future studies could try to relate certain leadership styles, such as transformational and transactional leadership [4][5], with a leader's observable behaviour in support of innovative behaviour.

*External demands related to client/customer relationships*: West [16] introduces external demands as a factor that influences innovative behaviour. In the author's findings, this influence was curvilinear (inverted U shape) meaning that too much or too little demand for innovation would have a negative effect on the expression of innovative behaviour. Within certain contexts in software development, client/customer participation in the development or the relationship with the software team could shape the external demand. We believe that this type of relationship should be investigated in software engineering, in particular in the case of agile development in which client participation tends to be more extensive than it is in traditional methods.

We used them to improve our next case study design, as part of our future work, as described by Monteiro et al. [14].

### 5.4 Lessons Learned
Before the pilot case study, we did not have established theory models to guide our investigation. We believe that the following lessons are important for researchers facing a similar situation:

*Dual role of the pilot case study*: a pilot study can support the development of provisional theories when none exists. It can also be instrumental in uncovering new factors or design issues not previously addressed.

*Case design of the pilot study:* do not use predefined models or theories that could prevent new variables from being uncovered, Keep the design simple, and collect as diverse a data set as possible within the constraints of the study.

*Use the results to refine the design*: after learning from the pilot study, one can decrease the breadth (variety) and increase the depth (focus) of data collection, focusing on relevant variables.

These lessons learned are detailed and further discussed by Monteiro et al. [14].

## 6. CONCLUSION
We presented the results of a pilot case study conducted to identify factors that influence the innovative behaviour of software engineers in practice. From these results, a preliminary model explaining the relationships among these factors was built (IBMSW-*i*), answering our research question. This result provides a rich description of factors that affect behaviour in different ways and the interaction among them, allowing a better understanding of the phenomenon in the software engineering practice. As far as we know, this is the first study to address this topic in software engineering practice.

Our model consistently integrates some of the previous proposal from the literature and extends these proposals with factors specific of the software development practice, such as the role of requirement stability. Further, we explicitly expressed individual attitudes as a key driver of innovative behaviour as well how feedback on previous innovative behaviour influences future expressions of this behaviour. These factors have not been addressed in the existing literature.

Being based on a pilot case study, we do not claim that the model is complete and universally generalizable (we would not claim generalizability for any case study), but it can be modified and extended with data and results from new contexts. Therefore, our model provides a well-founded starting point for future research, and in particular can help guide the design of other, full-scale case studies. Further, researchers and practitioners can use our model to learn the complexity of the studied phenomenon and to assess ways of transferring its findings to other contexts.

## 7. ACKNOWLEDGMENTS
Fabio da Silva holds a research grant from the Brazilian National Research Council (CNPq), process #314523/2009-0. Cleviton Monteiro received CAPES scholarship and the ELAP award.

## 8. REFERENCES
[1] Ajzen, I. 2012. The theory of planned behavior. In Lange, P., Kruglanski, A., Higgins, E. Handbook of theories of social psychology, Sage, London:UK, V.1, pp.438- 459

[2] Amabile, T., Conti, R., Coon, H., Lazenby, J., and Herron, M. 1996. Assessing the work environment for creativity. *Academy of Management Journal*, 39, 5, 1154-1184.

[3] Åmo, B. 2005. Employee innovation behavior. Bodø Graduate *School of Business*, Bodø:Norway

[4] Bass, B. M. Leadership and performance beyond expectations. Collier Macmillan, New York, 1985.

[5] Burns, J. 1987. Leadership. Harper & Row, New York, NY

[6] Charmaz, K.. Constructing Grounded Thoery: A Practical Guide Through Qualitative Analysis. Sage Publications, London, 2016.

[7] Christie, Michael, Rowe, Pat, Perry, Chad, and Chamard, John. 2000. Implementation of Realism in Case Study Research Methodology. In International Council for Small Business, Annual Conference (Brisbane)

[8] Cingöz, A. and Akdogan, A. 2011. An empirical examination of performance and image outcome expectation as determinants of innovative behavior in the workplace. Procedia - Social and Behavioral Sciences, 24, 847–853.

[9] Eisenhardt, K. 1989.Building Theories From Case Study Research.*The Academy of Management Review*,14,4,532-550

[10] França, A. C. C.; Sharp, h.; da Silva, F. Q. B. Motivated software engineers are engaged and focused, while satisfied ones are happy. In: Proceedings of the 8th ACM/IEEE International Symposium on Empirical Software Engineering and Measurement, 2014. doi>10.1145/2652524.2652545

[11] Jehn, K. A. et al. "Why Differences Make a Difference: A Field Study of Diversity, Conflict, and Performance in Workgroups", Administrative Science Quarterly, Vol. 44, 1999, pp. 741–763.

[12] Jong, J. P. J. and Den Hartog, D. N. 2007. How leaders influence employees' innovative behavior. *European Journal of Innovation Management*, 10, 41-64.

[13] Merriam, B. S. 2009. *Qualitative Research: A Guide to Design and Implementation*. Jossey-Bass, San Francisco.

[14] Cleviton V. F. Monteiro et Al. A Pilot Case Study on Innovative Behaviour: Lessons Learned and Directions for Future Work. In: Proceedings of the 10th ACM/IEEE International Symposium on Empirical Software Engineering and Measurement, 2016

[15] Scott, S.G. and Bruce, R.A. 1994. Determinants of innovative behavior: a path model of individual innovation in the workplace. *Academy of Management Journal*,38,1442-65.

[16] West, M.A. 2002 Sparkling fountains or stagnant ponds: an integrative model of creativity and innovation implementation in work groups. *Applied Psychology: An International Review*, 51, 3, 355-387